\magnification=\magstep1

\hsize=6.7truein
\hoffset=-0.20truein
\baselineskip=16truept plus 0 truept minus 0 truept
\overfullrule=0pt
\tolerance=10000

\newcount\notecnt
\notecnt=0
\def\adnote{\advance\notecnt by1}
\def\fnote{\adnote \footnote{$^{[\number\notecnt]}$}}

\pageno=1

\centerline{\bf Collision Rates in the Present-Day Kuiper Belt and Centaur Regions:}
\centerline{\bf Applications to Surface Activation and Modification}
\centerline{\bf On Comets, Kuiper Belt Objects, Centaurs, and Pluto-Charon}

\bigskip

\smallskip

\centerline{Daniel D.~Durda and S.~Alan Stern}
\centerline{Southwest Research Institute}
\centerline{Space Studies Department}
\centerline{1050 Walnut Street, Suite 426}
\centerline{Boulder, CO 80302}
\centerline{durda@boulder.swri.edu}

\centerline { }
\centerline { }
\centerline { }
\centerline { }
\centerline { }

\centerline { }
\centerline { }
\centerline { }

\noindent Submitted to {\it Icarus:} 26 March 1999

\noindent Revised: 19 November 1999
\bigskip

\noindent {\bf Length:}

20 Pages

08 Figures

00 Tables 

\smallskip

\vfill
\eject

\medskip
\noindent {\bf Proposed Running Header:} Collisions in the Edgeworth-Kuiper Belt
\medskip
\noindent {\bf Address editorial correspondence to:}
\medskip
\noindent Daniel D. Durda

\noindent Southwest Research Institute

\noindent 1050 Walnut Street

\noindent Suite 426

\noindent Boulder, CO 80302

\medskip

\noindent Tel: (303) 546-9670

\noindent Fax: (303) 546-9687

\noindent durda@boulder.swri.edu

\vfill
\eject

\medskip \centerline {\bf ABSTRACT}

\noindent \medskip We present results from our model of collision rates in the present-day
Edgeworth-Kuiper Belt and Centaur region.  We have updated previous results to allow for new
estimates of the total disk population, in order to examine surface activation and modification
time scales due to cratering impacts.  We extend previous results showing that the surfaces of
Edgeworth-Kuiper Belt objects are not primordial and have been moderately to heavily reworked by
collisions.  Objects smaller than about $r = 2.5$ km have collisional disruption lifetimes less
than 3.5 Gyr in the present-day collisional environment and have probably been heavily damaged
in their interiors by large collisions.  In the 30--50 AU region, impacts of 1 km radius comets
onto individual 100 km radius objects occur on $7 \times 10^7$--$4 \times 10^8$ yr time scales,
cratering the surfaces of the larger objects with $\sim$8--54 craters 6 km in diameter over a
3.5 Gyr period.  Collision time scales for impacts of 4 meter radius projectiles onto 1 km
radius comets range from 3--5 $ \times 10^7$ yr.  The cumulative fraction of the surface area of
1 and 100 km radius objects cratered by projectiles with radii larger than 4 m ranges from a few
to a few tens percent over 3.5 Gyr.   The flux of Edgeworth-Kuiper Belt
projectiles onto Pluto and Charon is also calculated and is found to be $\sim$3--5 times that of
previous estimates.  Our impact model is also applied to Centaur objects in the 5--30 AU region.
We find that during their dynamical lifetimes within the Centaur region, objects undergo very
little collisional evolution.  Therefore, the collisional/cratering histories of Centaurs are
dominated by the time spent in the Edgeworth-Kuiper Belt rather than the time spent on
planet-crossing orbits.  Further, we find that the predominant surface activity of Centaur
objects like Chiron is most likely not impact-induced.

\bigskip
\noindent {\bf Keywords:} Centaurs, Chiron, Comets, Kuiper Belt Objects, Pluto

\vfill
\eject

\medskip \noindent {\bf 1. INTRODUCTION}

\medskip Collisions are the dominant evolutionary process acting on most small bodies in the
solar system.  In the main asteroid belt, for instance, cratering collisions have greatly
modified the surfaces of individual asteroids by leaving large impact scars (e.g., Greenberg
{\it et al.}~1994, 1996; Veverka {\it et al.}~1997) and redistributing regolith across their
surfaces (Geissler {\it et al.}~1996), and catastrophic collisions over the aeons have left
their mark on the entire population size distribution (Davis {\it et al.}~1979, 1989; Durda {\it
et al.}~1998).

\medskip The Edgeworth-Kuiper Belt (EKB) population represents another major population of
small bodies whose evolution is largely shaped by collisions (Stern 1995).  Stern (1996) and
Davis and Farinella (1997) have further explored the rate of collisions between comets in the
region beyond 30 AU, and found that collisional evolution is a highly important process in
the EKB.  Collisional evolution in the EKB has recently been reviewed by Farinella
{\it et al.}~(2000).

\medskip Although intrinsic collision rates (number of collisions per kilometer$^2$ per
year) are lower by a factor of $\sim$1000 in the EKB compared to the main asteroid belt, the
population of objects there is $\sim$1000 times as great.  As a result of these competing
factors, the overall level of collisional processing of individual objects is of similar
scale to that in the main belt.

\medskip Here we seek to investigate the implications of the EKB collision rates for surface
modification.  In particular, we wish to estimate quantities such as the surface cratering
fractions, and the expected largest crater sizes.  In addition to a direct relevance for
understanding comets and other objects in the EKB, we also seek to gain insights into what the
Pluto-Kuiper Express spacecraft (Terrile {\it et al.}  1997) may observe when it images the
surfaces of Pluto, Charon, and other EKB objects.  Similarly, we seek to assess whether Centaur
objects on transient orbits in the giant planet region undergo further significant collisional
processing.

\medskip In what follows we first revisit previous collision rate calculations (Stern 1995,
1996) in light of both new observational data, and higher fidelity modeling.  Once improved
collision rates are obtained, we go on to evaluate their effect on the surfaces of objects in,
and derived from, the Edgeworth-Kuiper Belt.

\bigskip
\bigskip
\bigskip \noindent {\bf 2.  THE COLLISION RATE MODEL}

\medskip Stern (1995) examined collision rates in the present-day Edgeworth-Kuiper Belt
beyond 30 AU, as a function of the disk's radial and population size structure.  The numerical
model for calculating collision rates is described in detail in that paper, so only a brief
recapitulation will be presented in this section; in the next section we will describe changes and
improvements that have been made to the model to produce the results presented later in this paper.

\medskip The 1995 model is a static, multi-zone, multi-size-bin, particle-in-a-box collision rate
model that calculates instantaneous collision rates.  The colliding population is defined in
terms of a total disk mass and a single-valued power-law size distribution of objects in the
disk, normalized by the total number of $\sim$100 km diameter and larger objects in the
30--50 AU zone.  This size distribution is treated as a series of monotonically increasing
radius $r$ bins, with the objects in each successive bin 1.6 times larger in size (and 4
times more massive) than those in the preceding bin.\fnote{The radius of the smallest 
bin was 3.94 m; successive bin radii were 6.25 m, 9.92 m, etc.}  The model also
specifies the radial distribution of heliocentric surface mass density $\Sigma(r)$ so that:

$$
  \Sigma(r) = \Sigma_o r^{\beta}~{\rm ,} \eqno (1)
$$

\smallskip \noindent where $\Sigma_o$ is a normalization constant which in effect specifies a
total EKB mass in the 30--50 AU zone.  The power-law exponent $\beta$ determines the
heliocentric radial distribution of mass in the disk, with the two cases we consider defining
a realistic range of parameter space:  $\beta = -1$ corresponds to a constant mass per
heliocentric radial bin, while $\beta = -2$ (more realistic, and our preferred case),
corresponds to a declining mass per radial bin.  A disk-wide average eccentricity, $\langle e \rangle$, is adopted
for each model run; an equilibrium condition where the disk wedge angle $\langle i \rangle = {1 \over 2} \langle e
\rangle$ is assumed (see, e.g., Lissauer and Stewart 1993).

\medskip Once the global properties of the disk are specified, the disk is binned into a
series of radially concentric tori 1 AU in width, and the collision rates for objects at each
semimajor axis are then calculated in a particle-in-a-box formalism.  In this approach, the
instantaneous collision rate $\overline{c}$ (collisions/unit time) of target bodies with
semimajor axis $a$, eccentricity $e$, and radius $r_k$ being struck by impactors of radius
$r_l$ is

$$ 
\overline{c} (r_k,r_l,a,e,i,R) = 
$$
$$ \sum\limits_{R=a(1-\langle e \rangle )}^{a(1+\langle e 
\rangle )} \sqrt {{GM_{\odot}} \over {4 \pi^2 a^3}}~T(a,\langle e \rangle,R)
n(r_l,R)~v_{kl}(a,\langle e \rangle,\langle i \rangle,R)~\sigma_g(r_k,r_l,v_{kl},v_{esc[k+l]}), \eqno (2) 
$$

\smallskip \noindent where $T(a,\langle e \rangle,R)$ represents the time the target body
spends at each distance $R$ during its orbit.  $T(a,\langle e \rangle,R)$ is computed by
solving the Kepler time-of-flight equation explicitly for every $(a, \langle e \rangle)$ pair
in the model's parameter space.  The number density of impactors $n(r_l, R)$ in the torus
centered at distance $R$ is computed from the mass of the disk, the disk's wedge angle
$\langle i \rangle$, its population size distribution, and its heliocentric surface mass
density structure (Eq.~1).  Here $v_{kl}$ is the local average crossing speed of the impactor
population against the targets, $v_{\rm esc}$ is the escape speed of the combined
target-projectile pair, and $\sigma_g$ is the gravitational-focusing corrected
collision cross section.

\bigskip \noindent {\bf 3. MODEL IMPROVEMENTS AND INPUT PARAMETER UPDATES}

\medskip \noindent We have made two noteworthy improvements to the model outlined above.  These
are:

\medskip \item {$\bullet$} A more exact treatment of relative impact speeds.  In the earlier
model, relative impact speeds were calculated by a ``particle-in-a-box'' approximation of the
orbital motion of the target:  $v = (\langle e \rangle^2 + \langle i \rangle^2)^{1/2}v_k$,
where $v_k$ is the average Keplerian orbital speed of the target.  We now include in the
collision rate calculations the difference between the collision frequency of bodies in
mutual Keplerian orbits and that based on ``particle-in-a-box'' collisions, so that $v =
({5\over 4} \langle e \rangle^2 + \langle i \rangle^2)^{1/2}v_k$, as well as the effect of a
Gaussian speed distribution (cf., Wetherill and Stewart 1993, Appendix A).

\medskip \item {$\bullet$} Setting realistic limits on gravitational focusing.  Previously,
the effects of gravitational focusing were unconstrained, allowing the collision cross
section $\sigma_g$ to grow unrealistically large for the most massive targets and for very
low $\langle e \rangle$.  In the present model we now include limits on the gravitational
focusing factor due to Keplerian shear, 3-body effects, and velocity dispersion (cf., Ward
1996, Eqs.  9 and 11).

\medskip \noindent These improvements are numerical refinements affecting the final results at
only about the 10\% level as compared with our previous calculations; nonetheless, they are
worth documenting and make the final results more robust.  Of greater importance, we have
updated the input parameters necessary to compute collision rates in the EKB, based on
observational advances that have occurred since 1995.  In particular, these are:

\medskip \item {$\bullet$} Jewitt {\it et al.}~(1998) and Gladman {\it et al.}~(1998) have
each provided convincing evidence that between 30--50 AU there exist at least 70,000 objects
with $r>50$ km, and perhaps twice that many.  This is between 2 and almost 5 times the
population estimates for such bodies available in 1995.  We therefore conduct new model runs
with normalizations of both $7\times10^4$ and $1.4\times10^5$ objects with $r>50$ km.

\medskip \item {$\bullet$} Further, the population size distribution is now
represented by a more sophisticated, two-component power law of the
form $N(d_i) \propto d_i^{b}{\rm d}d_i$, where $b=-3$ for $d_i<d_0$ and $b=-4.5$ for
$d_i>d_0$, with $d_0=10$ km (Weissman and Levison 1997; hereafter WL97).\fnote{At large sizes the WL97 size distribution is consistent
with the latest estimates by other researchers (Gladman {\it et al.} 1999, for instance). For smaller, comet-size objects, simple,
single power-law extrapolations from larger sizes appear to over-estimate the
number of small EKB objects needed to supply the short-period comet flux (Duncan {\it et al.} 1995),
hence the broken power law of WL97.}

\medskip \noindent For reference, a WL97 size distribution, coupled with an estimated
population of 70,000 objects with $r > 50$ km, yields $\sim$$4564$ objects in our model's $r
= 102.4$ km size bin and $\sim$$1.2 \times 10^9$ objects in the $r = 1$ km size bin.  For a
population of 140,000 objects larger than $r = 50$ km, the number of objects in all size bins
doubles accordingly. We continue to model the spatial distribution of objects in
the 30--50 AU region as a disk, with our preferred surface mass density index $\beta = -2$, as
described above.

\medskip We have compared our modeled collision rates with those computed from the observed
distribution of EKO orbits (Bottke 1999; personal communication) and find very good agreement
between the two independent methods.  Our modeled collision rates, discussed in the following
sections, are within a factor of $\sim$2--4 of those calculated based on an \"Opik-style
collision rate model (Bottke {\it et al}.  1994) applied to the observed EKO orbit distribution.
Considering the fact that we have not made any attempt to bias-correct the observed orbit
distribution for this comparison, and the fact that our disk model has an inclination
distribution that is somewhat `colder' than the observed EKO population\fnote{Relative to
observed EKO eccentricities, observed inclinations are higher than the $\langle i \rangle = {1
\over 2} \langle e \rangle$ equilibrium values assumed in our model.}, we consider the agreement
between the two calculations quite good.\fnote {Nevertheless, we remind the reader about the
large model uncertainties attributable to using simple power laws for both the orbital
and the population size distributions.}

\bigskip \noindent {\bf 4. NEW ESTIMATES OF COLLISION RATES IN THE KUIPER BELT}

\medskip \noindent {\bf a. Collision Outcomes}

\medskip We now present results of collision rate calculations for the 30--50 AU
region obtained with our improved collision model and updated input parameters.

\medskip First, however, it is important to remember that given the dynamical conditions of the
present EKB, mutual collisions between Edgeworth-Kuiper Belt objects (EKOs) are generally
erosive.  That is, above some critical eccentricity, $e^*$, impacts occur at relative speeds
high enough that most ejecta escapes the target bodies.  Figure 1 shows the critical
eccentricity boundary between erosional (i.e., net mass loss) and accretional (i.e., net mass
gain) regimes for mutual collisions between EKOs (see Stern 1996).  Our contribution here, in
Figure 1, is the addition of some 128 multi-opposition EKOs for which fairly reliable orbits
have been determined, so that this large population of objects with moderately-well established
orbits can be evaluated relative to the critical eccentricity boundary curves.

\medskip Notice that the critical eccentricity for mutually colliding objects in the
EKB increases slightly with increasing heliocentric distance due to the direct linear
dependence of the typical approach speed upon the local Keplerian orbital speed.  Farther
from the Sun, higher $\langle e \rangle$'s are required to generate impact energies
sufficient to guarantee erosive collisions.  Thus, if $\langle e \rangle$ does not increase
with heliocentric distance, collisions will tend to be less erosional in nature as
we move outward through the EKB.  For  $\langle e \rangle$ greater than the critical
eccentricity, $e^*$, impacts eject more target mass than is retained, and the target
is either disrupted in response to a catastrophic collision, or eroded in the case of
a cratering collision.  

\medskip The plotted data points for 128 multi-opposition EKOs show that most large EKOs, like
the main-belt asteroids, are currently undergoing predominantly erosive collisions, even under
the most pessimistic assumption, i.e., that of strong surface mechanical properties.  As to
classical, km-scale comets (i.e., those objects which leave the EKB to appear as the Jupiter
Family comets), $e^*$ values are so low as to guarantee that these bodies have resided in a
heavily erosional collisional environment.

\medskip \noindent {\bf b. Collision Rates and Fluxes}

\medskip Figure 2 shows the typical number of collisions which occur onto 100 km- and 1 km-scale
radius EKOs as a function of projectile radius, at both 35 and 45 AU.

\medskip The calculations shown here assume the ``nominal'' estimated population\fnote{In what
follows we will report collision time scales and impact fluxes for the ``nominal'' population
only.  For a larger population with $N(r>50~{\rm km}) = 1.4 \times 10^5$, reported collision
time scales for individual objects will be reduced by a factor of~2 and impact fluxes will be
increased by a factor of~2.}  of such objects in the 30--50 AU zone today, i.e., $N(r>50~{\rm
km}) = 7 \times 10^4$.  Since the collision rate model used here is static (i.e., it calculates
collision rates for the present disk and does not account for a decrease in the population size
with time as bodies are collisionally destroyed), we calculate the total flux of impactors on a
target only over the last 3.5 Gyr, the approximate time since which the disk is expected to have
reached roughly its present mass and dynamical state (e.g., Weissman and Levison 1997).\fnote
{For this reason, we denote any object or surface unit on an object of this age as
``primordial'' if it is this age or older.}  Our interest here is in present-day collisional
rates and effects.  This is primarily because the small bodies in the EKB are young compared to
3.5 Gyr (Stern 1995; Davis and Farinella 1997), and bodies of all sizes underwent far more
significant collisional processing in the more massive, primordial Kuiper disk (Stern and
Colwell 1997).

\medskip Figure 2 shows that a typical, 100 km-scale radius EKO will have undergone $\sim$8--54
cratering impacts with 1 km radius ``comets'' over the last 3.5 Gyr, depending on heliocentric
distance and $\langle e \rangle$.  At 35 AU the collision time scales for a typical EKO are $6.5
\times 10^7$ and $1.5 \times 10^8$ yr for $\langle e \rangle = 0.0256$ and 0.2048, respectively.
The same collision time scales are $1.4 \times 10^8$ and $4.1 \times 10^8$ yr at 45 AU for the
same $\langle e \rangle$'s.  These values of $\langle e \rangle$ cover the range of observed $e$
for most EKOs.  For large EKOs, collision time scales are shorter for smaller $\langle e
\rangle$'s due to increased gravitational focusing effects at smaller encounter speeds.  Given
our estimated population of 100 km radius EKOs in the EKB (4564 objects in the size bin with
radii between 81--129 km), there should be one such EKO-comet collision somewhere in the 30--50
AU region every $\sim$$1.4$--$9.0 \times 10^4$ yr.  Smaller projectiles hit more frequently,
with impact time scales for 4 m radius projectiles onto any single 100 km-scale radius target of
$\sim$1000--6000 yrs; across the entire EKB, such cratering impacts occur every $\sim$80--510
days.

\medskip Figure 2 also shows the number of collisions onto a 1 km radius comet.  Over 3.5
Gyr\fnote{In the next subsection we show that catastrophic disruption lifetimes for
1 km radius comets in the EKB range from $\sim$1--10 Gyr. We calculate impact fluxes here
over 3.5 Gyr in order to directly compare with results for larger objects; for shorter- or
longer-lived comets, fluxes will be proportionately smaller or larger, respectively.} a 1~km radius comet between 35 and 45 AU will experience $\sim$90--300 cratering
collisions with projectiles larger than $r = 4$ m.  Collision time scales for impacts of 4 m
radius projectiles onto 1 km radius targets range from 2.5--$4.7 \times 10^{7}$ yr.  Over the
entire population of $\sim$$2 \times 10^{9}$ comets in the disk, there should be one such
collision every several days. These numbers assume, of course, that the WL97 size
distribution is valid to sizes as small as 4 m. The actual size distribution of
small objects in the EKB is highly uncertain, and the cratering records of the Galilean
satellites may even hint at a lack of cometary objects smaller than $\sim$100 m (Chapman 1997; Chapman {\it et al.} 1998).

\bigskip \noindent {\bf c. Catastrophic Disruption Lifetimes and Deep Interior Modification}

\medskip What about larger collisions and the likelihood that a 100 km radius EKO would be
catastrophically disrupted over 3.5 Gyr?  The impact scaling literature has been dominated by
consideration of collisions between asteroids, so the impact specific energies and scaling
laws used here are most appropriate for silicate targets.  However, Ryan {\it et al.}  (1999)
have conducted laboratory impact studies in which porous ice targets were impacted by
fractured ice projectiles to simulate collisions between EKOs, and found that impact specific
energies and fragmentation modes are similar to those for silicate targets.  Smooth
particle hydrodynamics calculations by Benz and Asphaug (1999) indicate that impact specific energies for basalt targets are only $\sim$2--4 times greater than those for icy objects under the same impact conditions. Estimates of the
impact specific energy, $Q^*_D$, required to catastrophically disrupt and disperse a 100 km
radius object range from $\sim$1--4$ \times 10^5$ J kg$^{-1}$ (Davis {\it et al.}  1989; Love
and Ahrens 1996; Melosh and Ryan 1997; Durda {\it et al.}  1998).

\medskip With $\langle e \rangle = 0.2048$, effective relative impact speeds between 100
km radius EKOs and other objects in the 30--50 AU region range from $\sim$1.1--1.4 km
s$^{-1}$.  The relative impact speed is $v_i = (U^2 + v_{\rm esc}^2)^{1/2}$, where $U$ is
the hyperbolic encounter speed (dependent upon heliocentric distance and $\langle e \rangle$\fnote{The encounter speed is also dependent upon $\langle i \rangle$, but
recall that we have assumed an equilibrium condition where $\langle i \rangle = {1 \over 2} \langle e\rangle$.})
and $v_{\rm esc}$ is the escape speed of the combined target-projectile pair.  At these
speeds, the smallest object capable of delivering the required specific energy for disruption
to the target is about $r =$ 53--84 km in size.  The time scale for such a collision is
presently 3--8$ \times 10^{12}$ yr (using a model bin radius of 64.5 km). Still-larger
EKOs would have longer disruption time scales.

\medskip Our results show clearly that the vast majority of the largest EKOs are not likely to have been involved
in disruptive collisions over the last 3.5 Gyr.  This is in agreement with collisional models
of EKB evolution by Davis and Farinella (1997), showing that the population at diameters
larger than about 100 km is essentially unchanged over solar system history.

\medskip Figure 3 shows the catastrophic disruption lifetimes for a range of EKO sizes at 35 and
45 AU with $\langle e \rangle = 0.2048$.  Shattering time scales for 100 km radius objects are
shorter, $\sim$3.9--10.5 Gyr, since projectiles as small as $r \approx 5$ km may fragment the
target without dispersing the resulting debris, resulting in gravitationally re-accumulated
rubble-pile EKOs.

\medskip What is the largest object size-class in which the majority of objects can be expected to have been involved in disruptive collisions
in the EKB in the last 3.5 Gyr?  Using our highest modeled $\langle e \rangle$ of 0.2048 and
the lowest estimated $Q^*_D$ for objects of various sizes, and searching through the many
combinations of our calculated target-projectile collision time scales, we find that objects
smaller than $r \approx 2.5$ km presently have catastrophic disruption time scales at 35 AU
less than 3.5 Gyr.  At 45 AU the disruption time scale increases by a factor of $\sim$2.5--3,
so that smaller objects with radii less than about 1.6 km have lifetimes of 3.5 Gyr.  At
lower $\langle e \rangle$ and assuming values of $Q^*_D$ nearer the mid-range of published
scaling laws, objects become harder to destroy, so that $r \approx 2.5$ km should be
considered an upper limit to the size object that can have a collisional disruption time scale
less than 3.5 Gyr anywhere in the EKB.

\medskip From the collision rates yielded by our model, we can also calculate the collisional
disruption lifetime in the EKB for a typical comet nucleus with radius $r = 1$ km.  We
estimate the impact specific energy for disrupting such an object to be $\sim$10--200 J
kg$^{-1}$ (Durda {\it et al.}  1998; Melosh and Ryan 1997).  Using these values for $Q^*_D$,
and assuming disk-wide $\langle e \rangle$ of 0.0256--0.2048, we find that at 35 AU a $r=1$
km comet can be destroyed by $r \approx$ 0.02--0.25 km projectiles.  The corresponding
disruption lifetime ranges from $\sim$$9.6 \times 10^{8}$--$10^{10}$ yr.  At 45 AU the
lifetime is $\sim$$2.6 \times 10^{9}$--$10^{11}$ yr.  Eccentricities for most observed EKOs
tend to be near the higher of the values assumed here, so that disruption time scales for
comets are likely near the lower values reported.  Across the 30--50 AU region, we find collision
lifetimes for $r = 1$ km comets are then likely of order 1--10 Gyr, assuming mid-range values
of $Q^*_D$.

\bigskip
\bigskip
\bigskip \noindent {\bf 5. SURFACE MODIFICATION IN THE EKB}

\medskip \noindent {\bf a. Cratering Fraction}

\medskip The results of collision rate calculations presented in the previous section
demonstrate that EKOs of all sizes have suffered a significant number of collisions.  These
collisions can be expected to have significantly affected the surfaces of both large and
small objects, covering their surfaces with craters of various sizes, overturning and
reworking underlying, more primordial surfaces, removing surface materials through impact
sublimation, and possibly exposing deeper and more volatile icy species and thus possibly
serving as a mechanism to activate distant, inactive objects.

\medskip Using the encounter speeds and impactor fluxes calculated in the previous section,
we can estimate the sizes and spatial coverage of impact craters on typical EKOs.  Holsapple
(1993) gives an expression for the diameter of an idealized, hemispherical crater, as:

$$
  D = 1.26 d (A\rho_{\rm impactor}/\rho_{\rm target})^{1/3}(1.61 g d/v_i^{2})^{-\alpha/3}~{\rm ,} \eqno (3)
$$

\noindent where $d$ is the impactor diameter, $\rho$ is the density, $g$ is the surface
gravity of the target, $v_i$ is the encounter speed, and $A$ and $\alpha$ are constants
dependent upon the mechanical properties of the target material.  Holsapple (1993) uses
values of $A$ and $\alpha$ of 0.2 and 0.65, respectively, for water ice.  For a 100 km-scale radius
EKO with an assumed density of $\rho = 1.5$ g cm$^{-3}$, $g \approx 4.3$ cm sec$^{-2}$.  The
relative encounter speed, $v_i$, as defined in the previous section, includes the escape
speed of the combined target-projectile pair.

\medskip Figure 4 shows the fraction of a $r = 102$ km target's surface area covered by
craters produced by projectiles of various sizes, over a 3.5 Gyr period.  We find
that the cumulative fraction of the surface cratered by all $r > 4$ m projectiles ranges from
$\sim$7--32\%, for targets from 35--45 AU and for $\langle e \rangle =$ 0.0256--0.2048.  This
is a conservative estimate that does not include the additional surface area covered by
crater flanks and ejecta blankets, which would cover $\sim$4 times more area.  Significant
additional area will also be covered by craters produced by projectiles smaller than $r = 4$ m,
our lower size limit in these calculations. Overall, perhaps a third of the entire surface of
a typical 100 km-scale EKO will have been re-worked by impact cratering over the past
3.5 Gyr.

\medskip Collision rates calculated in the previous section showed that comet-sized objects ($r
\approx 1$ km) will also be significantly cratered during their lifetimes, typically enduring
$\sim$90--300 impacts by objects larger than $r = 4$ m.  Figure 4 shows that a typical comet
with $r=1$ km will have $\sim$20--224\% of its surface cratered by impacts with $r>4$ m
projectiles over 3.5 Gyr.  Davis and Farinella (1997) concluded that most comet-size objects are
collisional fragments of larger parent objects.  Our results suggest that even the small
fraction of objects surviving, ``original'' accretion products objects (i.e., those that are
$>$3.5 Gyr old) will have undergone substantial collisional processing in the form of cratering
and sub-catastrophic impacts.

\medskip This result extends the notion that the surfaces of EKOs are not primordial and have
been moderately to heavily reworked during their history in the EKB (Stern 1995; Luu and Jewitt
1996; Davis and Farinella 1997; Farinella {\it et al.}  2000).  The numerous smaller impactors,
responsible for most of the spatial coverage, will have comminuted and gardened the surfaces of
EKOs, producing a regolith of shattered, icy debris.  As we will see below, EKO surfaces are
expected to display both fresh and primordial (i.e., $>$3.5 Gyr old) units.

\medskip Larger impactors will have penetrated deeper into the target's surface, excavating
material that is less impact-processed from below.  The most recent, large impacts may have
associated bright ejecta blankets and ray systems.  Brown {\it et al.} (1999) report
evidence that the intensity of water bands in the spectrum of 1996 TO$_{66}$ varies
with rotational phase, which suggests a ``patchy'' surface. Fig.  2 shows that the largest impactors
likely to have struck 100 km radius EKOs over the last 3.5 Gyr have $r \approx 5$ km, resulting
in craters $\sim$26 km in diameter.  Deeper, less impact-processed material may contain more
volatile ice species that could provide a source of surface activity around recent impact
sites.  Such activity may range from small, geyser-like plumes activated by impactors just
large enough to penetrate a processed regolith, up to the production of full-scale comae
generated by the largest impactors (see Fitzsimmons and Fletcher [1999] in this regard).

\bigskip \noindent {\bf b. Mass Loss Due to Ejecta Escape}

\medskip The same impacts responsible for overturning and reworking the surface of an EKO and
covering its surface with craters will also erode and remove a fraction of that surface
through the escape of ejecta launched at speeds greater than the escape speed of the
target.

\medskip Standard impact scaling laws indicate that for strong targets with radii less than
$\sim$75--150 m, ejecta speeds should exceed the escape speed of the object.  In a purely
strength-scaling cratering regime, the volume eroded from the target is directly proportional to
the volume of the impactor:  $V_{erode}(D) = hD^3$, where $D$ is the projectile diameter, and $h
\approx 120$ as suggested by Holsapple (1993).  For larger targets, such as a 100 km-scale
radius EKO, larger craters are excavated in the gravity-scaling regime and a much smaller
fraction of the ejecta is launched at greater than the escape speed.  Even for targets as small
as 1 km radius comets, a significant fraction of impact ejecta may be retained (Ryan and Melosh
1998).  For a 100 km-scale radius icy EKO, and the mechanical properties described
just below Eqn.~(3), the transition crater diameter between the
strength-scaling and gravity-scaling regimes is of order 40--50 m, about the size of craters
produced by the smallest objects considered in our model ($r$=4 m).  As in the
strength-scaling regime, the volume eroded per volume of projectile for gravity-scaling
cratering is independent of the size of the impactor, $V_{erode}(D) = h^{\prime}D^3$, but here
$h^{\prime}$ is much smaller than $h$, of order 3--4 for a 100 km-scale radius icy object in the
EKB (see the Appendix of Geissler {\it et al.}  (1996) for a more
complete discussion of mass loss from small objects due to impact ejecta erosion).

\medskip With the above mass loss assumptions and the EKO population and collision time scales
calculated from our model, we can estimate the magnitude of the surface loss from comets and
large EKOs due to escape of impact ejecta.  Time-averaged mass loss rates from a 100 km radius
EKO due to impacts with $r > 4$ m projectiles range from $\sim$$4 \times 10^8$ to $7 \times
10^9$ g yr$^{-1}$ for $R$ between 35 and 45 AU and $\langle e \rangle$ between 0.0256 and
0.2048.  Over 3.5 Gyr, this amounts to some 10--100 m of surface loss.  Mass loss rates from 1
km radius comets range from $6 \times 10^5$ to $2 \times 10^6$ g yr$^{-1}$ under the end-member
assumption of cratering in the strength regime, amounting to $\sim$110--390 m of surface loss
over 3.5 Gyr.  Gravity-regime cratering would result in lower mass loss rates and less surface
erosion due to the larger fraction of retained ejecta.  We suspect that most of the surface loss
effects due to impacts will be localized primarily around larger impact sites, and the
impact-undisturbed surface fraction (perhaps 90\% for a minimum impact radius of 4 m) will be
considerably less damaged.  In contrast, recall that fully a third of the entire surface of a
typical large EKO may have been processed over the past 3.5 Gyr by mechanical overturning and
regolith formation.

\vfill \eject

\noindent {\bf 6. COLLISION RATES ON PLUTO AND CHARON RE-EVALUATED}

\medskip Weissman and Stern (1994) estimated impact rates of EKB and Oort Cloud
comets onto both Pluto and Charon given early estimates of EKB population numbers.  They showed that the
outer Oort Cloud is a negligible source of impactors and that the dominant sources of
impactors on both bodies are EKB and inner Oort Cloud comets, with the EKB population dominating over the inner OC by factors of a few,
contributing a flux of some 2400 impacts over the age of the solar system onto Pluto's
surface, and 460 impacts onto Charon's surface.  Here we re-examine the EKB flux onto Pluto
and Charon using newer EKB population estimates described above in \S3.

\medskip Figure 6 shows the number of EKO impacts onto Pluto and Charon over 3.5 Gyr, as a
function of impactor radius.  As above, the calculated flux of impactors includes impact
cross-section enhancement due to the focusing effects of Pluto's and Charon's gravitational
fields.  Over 3.5 Gyr, the total number of $r = 1$ km comets striking Pluto and Charon is
approximately $8.9 \times 10^3$ and $1.1 \times 10^3$, respectively.  (To compare with the
Weissman and Stern (1994) results, this amounts to some $1 \times 10^4$ and $1.2 \times 10^3$
impacts onto Pluto and Charon, respectively, over 4.5 Gyr).  Clearly, improved EKB population
parameters have increased the Weissman and Stern (1994) fluxes by $\sim$3--5 times.  Impacts of
1~km radius comets onto Pluto occur on time scales of $\sim$$3.9 \times 10^5$ yr.  Similar
impacts on Charon occur on $\sim$$3.2 \times 10^6$ yr time scales.  The largest EKOs expected to
have impacted Pluto and Charon during the last 3.5 Gyr have radii of $r=40$ and 20 km,
respectively.  We find from Eq.  3 that the resulting largest crater diameters on both bodies
due to present-day EKB collisions are roughly 123 and 75 km, respectively.  Larger impact basins
resulting from earlier, massive impacts may well underly these more recent craters.

\medskip If we simply sum up the total area covered by craters on Pluto's surface over 3.5
Gyr, we find cumulative cratered surface fractions of $\sim$40\%, 0.4\%, and 0.003\% for $r >
4$, 40, and 400~m projectiles, respectively.  However, the hydrodynamic escape of Pluto's
atmosphere (Trafton {\it et al.}  1997) implies that there has been $\sim$1--5 km of surface
loss due to sublimation, if the present escape flux has been maintained over the age of the
solar system.  For craters and basins with depths of 1 km ($\sim$5 km diameter), Pluto's surface may
therefore be comparatively young, of order $2 \times 10^8$ yr.  On such a time scale, the
cumulative cratered fraction drops to $\sim$2\%, 0.02\%, and 2$\times$10$^{-4}$\% for $r >
4$, 40, and 400~m projectiles, respectively.

\medskip Unlike Pluto, Charon's surface is not losing significant volatiles to atmospheric
escape (Trafton et al.~1997), and so should record the cumulative flux of projectiles
encountered over its lifetime.  Over 3.5 Gyr, the cumulative cratered fractions for Charon are
$\sim$20\%, 0.2\%, and 0.002\% for $r > 4$, 40, and 400~m projectiles, respectively.

\medskip When the highly-anticipated Pluto-Kuiper Express reconnaissance mission reaches the
Pluto-Charon system, one should quite clearly expect Charon to display an older surface
reflecting the time-integrated flux of projectiles.  At the same time, Pluto's surface should
reflect the recent production population, thereby showing the essentially-instantaneous
(i.e., recent-times) impact flux.

\medskip Given the modeled EKB flux through the Pluto-Charon system, we can also calculate the
size of the smallest surviving primordial satellite in the system.  The Pluto-Charon pair likely
formed as the result of a giant impact which may have left smaller satellites or debris orbiting
Pluto after the accretion of Charon.  We calculate that at 39 AU, with $\langle e \rangle =
0.2048$, objects smaller than $r \approx 1.5$--2 km have catastrophic disruption lifetimes less
than 3.5 Gyr.  Any primordial Pluto satellites smaller than this should have been destroyed by
collisions with EKB projectiles.  Analysis of archival HST images by Stern {\it et al.}  (1994)
shows that at the 90\% confidence level, no Pluto satellites larger than $r \approx 140$, 46,
and 42 km exist inside the Charon instability strip, between 1 and 2 arcsec from Pluto (i.e.,
between 1.1 and 2.2 Charon's orbital radius), and outside 2 arcsec from Pluto, respectively.  If
the Pluto-Charon system harbors undiscovered, surviving primordial satellites, we conclude that
they will most likely be in the $r \approx 2$--46 km size range.

\bigskip \noindent {\bf 7.  COLLISION RATES IN THE CENTAUR REGION}

\medskip We have also applied our model to examine collision rates in the Centaur region,
between 5 and 30 AU.  Slow leakage of objects from the EKB due to planetary perturbations
sustains a population of objects in the giant planet region with dynamical lifetimes of order $5
\times 10^7$ yr (Duncan {\it et al.}  1995; Levison \& Duncan 1997).  Chiron, Pholus, and the
other presently known Centaurs in this region are thus recognized as emissaries from the EKB,
delivered to a less distant region of the solar system, enabling more detailed observational
studies of Centaurs than of EKOs.  We therefore wish to understand how EKB collisional histories
recorded on the surfaces of Centaurs have been modified in the Centaur region.

\medskip Figure 7 shows a collisional erosion/accretion boundary plot similar to Fig.~1, but
for the population of known Centaurs.  The higher average orbital speeds and eccentricities of
the Centaurs conspire together to place them strongly in the erosive regime, so that growth
of larger objects by accretion is not possible in that region at the present time.  Such
erosive collisions will both contribute ejecta and fine dust in the 5--30 AU zone, and cause
EKOs in that region to slowly, but surely, lose mass while in transit through this region.

\medskip To quantify such effects, we constructed a model disk of Centaur objects based upon
observational constraints on the population by Jedicke and Herron (1997).  Their
determination of the detection efficiency of the Spacewatch survey system, combined with the
observed absolute magnitude distribution of observed Centaurs and a model of the orbit distribution based on numerical integrations of Levison and Duncan (1997), allowed Jedicke and Herron to
determine that in the 5--30 AU region there must be fewer than $\sim$2000 objects in the
absolute magnitude range $-4 < H < 10.5$ ($r > 26$ km for $p_{\nu} = 0.04$).  Assuming that
the Centaur absolute magnitude distribution can be represented by a power law, they find a
slope parameter $b = 4.05$ over the relevant size range.  This slope is just slightly less
steep than that favored for EKOs by Weissman and Levison (1997) over the same size range.  As
the Centaurs are a dynamical sampling of the EKB population, we therefore continue to use the
WL97 population size distribution as our favored size distribution, but we will also report
Centaur collision rates assuming the Jedicke and Herron best-fit power law.  Finally, we
distribute the estimated total number of Centaurs throughout a disk ranging from 5--30 AU
with a heliocentric surface mass density dependence of $R^{1.3}$ (Levison and Duncan 1997),
and calculate collision rates as above.

\medskip Figure 8 shows the resulting number of impacts onto a Chiron-size object ($r \approx
90$ km) at 14 AU, over a typical $5 \times 10^7$ yr dynamical lifetime, as a function of
impactor size.  Collision time scales at 14 AU are $\sim$300 times longer than for a
comparable body at 40 AU.  At 14 AU, cratering collisions of 1 km radius comets onto
Chiron-size targets occur every $\sim$60 Gyr (every $\sim$4 Gyr assuming the Jedicke and
Herron best-fit size distribution extends to $r = 1$ km).  Clearly, there is
very little collisional evolution in Centaur region.  We therefore conclude that the cumulative collisional and
cratering history of Centaurs is dominated by the time they spent in Kuiper Belt, and that only a
negligible amount of collisional processing occurs while they are extant in the Centaur region.

\medskip Hughes (1991) has proposed that surface activity responsible for outbursts and coma
around Chiron might be the result of cratering impacts exposing supervolatile ices.  Analysis
of pre-discovery images of Chiron (Bus {\it et al.}~1999) indicates variability in surface activity
on time scales of a few to several years.  Our model calculations for Chiron yield cratering
time scales with 4 m radius projectiles of $\sim$1 Myr for a WL97 size distribution.  Even
assuming that the Jedicke and Herron power law size distribution is valid to sizes as small
as 4 m, the mean
time between impacts will be of order 200 yr, significantly longer than the observed
time scale of outburst activity on Chiron.  We might be fortunate to witness one such
cratering impact on a single object, but given the lengthy time scales, one must conclude that
the predominant surface activity on Chiron is not caused by impacts.

\bigskip \noindent {\bf 8. CONCLUSIONS AND DISCUSSION}

\medskip We have updated our previous collision rate model (Stern 1995) to include a more
precise treatment of encounter speeds and collision cross sections, and have incorporated new
estimates of the EKB population size and structure (WL97). Based on this we find:

\medskip 1.  Collision time scales in the present EKB for 1 km radius comets onto 100 km
radius objects are $\sim$$6.5 \times 10^7$--$4.1 \times 10^8$ yr.  Over 3.5 Gyr this amounts
to $\sim$8--54 such impacts onto a single 100 km target.  Given the estimated population of
such objects in the present EKB, there should be one such impact somewhere in the 30--50 AU
region every $\sim$1.4--$9.0 \times 10^4$ yr.  Impacts of 4 m radius projectiles onto 1 km
radius comets occur on 2.5--$4.7 \times 10^7$ yr time scales, resulting in $\sim$90--300
cratering impacts with $r > 4$ m projectiles onto individual comets.  Over the entire
population of $\sim$$2 \times 10^9$ comets in the EKB, there should be one such collision
every few days.

\medskip 2.  Assuming relative encounter speeds of $\sim$1.1--1.4 km s$^{-1}$ between objects
in the 30--50 AU region, and using impact strengths from published scaling laws, we estimate
that 100 km-scale radius EKOs can be catastrophically disrupted by 53--84 km radius projectiles,
yielding disruption lifetimes of 3--$8 \times 10^{12}$ yr in the present EKB.  Catastrophic
disruption time scales for 1 km radius comets range from 1--10 Gyr.

\medskip 3. Objects smaller than about $r = 2.5$ km have collisional disruption lifetimes
less than 3.5 Gyr in the present-day EKB collisional environment. It can be expected that most small, comet-size bodies,
even primordial objects not formed as collision fragments, have been heavily damaged in their interiors
by large collisions.

\medskip 4.  The cumulative fraction of the surface area of 1 and 100 km radius objects
cratered by projectiles with $r$$>$4 m ranges from a few to a few tens percent
over 3.5 Gyr. 

\medskip 5.  Over 3.5 Gyr, Pluto and Charon are estimated to have been impacted by $8.9
\times 10^3$ and $1.1 \times 10^3$ 1~km radius comets, respectively.  Impacts of 1~km radius
comets onto Pluto occur on time scales of $\sim$$3.9 \times 10^5$ yr.  Similar impacts on
Charon occur on $\sim$$3.2 \times 10^6$ yr time scales.  Because of the hydrodynamic escape of
Pluto's atmosphere, its surface may be comparatively young for craters with depths less than
about 1 km.  In this case, fresh craters may cover less than 2\% of Pluto's surface
due to impacts by projectiles with radii greater than 4 m.  Charon's surface, in contrast,
should appear substantially older, recording a history of impacts since its formation, and
having more than 20\% of its surface cratered by projectiles with radii greater than 4 m.

\medskip 6.  In the Centaur region, collisions of 1 km radius comets onto 100 km radius
targets (roughly Chiron's size) occur on time scales as long as $\sim$60 Gyr.  Collision time scales at
14 AU are $\sim$300 times longer than those for comparable bodies at 40 AU.  The collisional and
cratering histories of Centaurs are dominated by the time they spent in the EKB. The predominant surface activity on Chiron is not likely caused by impacts.

\medskip Our results, like those of Davis and Farinella (1997) and Stern (1995), show that the small bodies we
call comets coming from the EKB must be young compared to the age of the Solar System.  This
will not be the case in the Oort Cloud, where Stern (1988) showed collisions are rare.  Thus,
we can predict a major difference between OC and EKB comets:  age, which should manifest
itself in CRE age, crater counts, regolith reworking, radiation effects on surface
microstructure, and perhaps even albedo and upper crust chemistry.

\medskip 

\centerline { }
\centerline {\bf Acknowledgements}

\medskip We thank Eileen Ryan and Stan Love for helpful reviews.  We also thank Robin Canup
and Joel Parker for their comments on our draft manuscript, Hal Levison for helpful discussions,
and Bill Bottke for his independent calculations of collision rates as a check for our model.
This work was supported by the NASA Origins of Solar Systems Program.

\vfill
\eject

\centerline {\bf References}

\medskip \noindent Ahrens, T. J., and S. G. Love 1996. Strength versus gravity dominance
in catastrophic collisions. {\it Lunar Planet. Sci.} {\bf 27}, 1--2.

\medskip \noindent Bottke, W. F., M. C.~Nolan, R.~Greenberg, and R.~Kolvoord 1994. 
Velocity distributions among colliding asteroids. {\it Icarus} {\bf 107}, 255--268.

\medskip \noindent Benz, W., and E. Asphaug 1999. Catastrophic collisions
revisited. Submitted to {\it Icarus}.

\medskip \noindent Brown, R. H., D. P.~Cruikshank, and Y.~Pendleton 1999. Water
ice on Kuiper Belt object 1996 TO$_{66}$. {\it Astrophys. J.} {\bf 519}, L101--L104.

\medskip \noindent Bus, S. J., M. F.~A'Hearn, E.~Bowell, and S. A.~Stern 1999.
2060 Chiron: Evidence for Activity Near Aphelion. Submitted to {\it Icarus}.

\medskip \noindent Bus, S. J. 1991.  Detection of CN emission from (2060)~Chiron.  {\it
Science} {\bf 251}, 774--777.

\medskip \noindent Chapman, C. R. 1997. Cratering on the Galilean satellites:
Implications for the size distribution of cratering impacts in the solar
system.
{\it Meteoritics and Planetary Science} {\bf 32}, A27.

\medskip \noindent Chapman, C. R., W. J.~Merline, B.~Bierhaus, S.~Brooks, and
the Galileo Imaging Team 1998. Cratering in the Jovian system: Intersatellite
comparisons. {\it Lunar Planet. Sci.} {\bf 29}, abstract no. 1927.

\medskip \noindent Davis, D. R., C. R.~Chapman, R.~Greenberg, and A. W.~Harris 1979.
Collisional evolution of asteroids: Populations, rotations, and
velocities. In {\it Asteroids} (T.~Gehrels, Ed.), pp. 528--557. Univ.~of
Arizona Press, Tucson.

\medskip \noindent Davis, D. R., E. V.~Ryan, and P.~Farinella 1995. On how to
scale disruptive collisions. {\it Lunar Planet. Sci.} {\bf 26}, 319--320.

\medskip \noindent Davis, D. R., and P.~Farinella 1997. Collisional evolution
of Edgeworth-Kuiper belt objects. {\it Icarus} {\bf 125}, 50--60.

\medskip \noindent Davis, D. R., S. J.~Weidenschilling, P.~Farinella, P.
~Paolicchi, and R. P.~Binzel 1989. Asteroid collisional history: Effects on
sizes and spins. In {\it Asteroids II} (R. P.~Binzel, T.~Gehrels, and M. S.
~Matthews, Eds.), pp. 805--826. Univ. of Arizona Press, Tucson.

\medskip \noindent Duncan, M. J., H. F.~Levison, and S. M.~Budd 1995. The dynamical
structure of the Kuiper Belt. {\it Astron. J.} {\bf 110}, 3073--3081.

\medskip \noindent Durda, D. D., R.~Greenberg, and R.~Jedicke 1998. Collisional
models and scaling laws: A new interpretation of the shape of the main-belt
asteroid size distribution. {\it Icarus} {\bf 135}, 431--440.

\medskip \noindent Farinella, P., D. R.~Davis, and S. A.~Stern 2000. Formation and
collisional evolution of the Edgeworth-Kuiper Belt.  In {\it Protostars and Planets IV} 
(V.~Mannings and S.~Russel, Eds.), Cambridge University Press. In press.

\medskip \noindent Geissler, P., J.-M.~Petit, D. D.~Durda, R.~Greenberg, W.~Bottke,
M.~Nolan, and J.~Moore 1996. Erosion and ejecta reaccretion on 243 Ida and its
moon. {\it Icarus} {\bf 120}, 140--157.

\medskip \noindent Gladman, B., J. J.~Kavelaars, P. D.~Nicholson, T. J.~Loredo,
and J. A.~Burns 1998. Pencil-beam surveys for faint trans-Neptunian comets. {\it Astron. J.} {\bf 116}, 2042--2054.

\medskip \noindent Greenberg, R., W.~Bottke, M.~Nolan, P.~Geissler, J.-M.~Petit,
D. D.~Durda, E.~Asphaug, and J.~Head 1996. Collisional and dynamical history of
Ida. {\it Icarus} {\bf 120}, 106--118.

\medskip \noindent Greenberg, R., M.~Nolan, W. F.~Bottke, Jr., R. A.~Kolvoord,
and J.~Veverka 1994. Collisional history of Gaspra. {\it Icarus} {\bf 107},
84--97.

\medskip \noindent Holsapple, K. A. 1993. The scaling of impact processes in
planetary sciences. {\it Ann. Rev. Earth Planet. Sci.} {\bf 21}, 333--374.

\medskip \noindent Hughes, D. W. 1991. Possible mechanisms for cometary outbursts.
In {\it Comets in the Post-Halley Era} (R. L. Newburn, M. Neugebauer, and
J. Rahe, eds.). Kluwer, Dordrecht, 825--854.

\medskip \noindent Jedicke, R., and J. D.~Herron 1997. Observational constraints
on the Centaur population. {\it Icarus} {\bf 127}, 494--507.

\medskip \noindent Jewitt, D., and J. X.~Luu 1993.  Discovery of Kuiper Belt Object 1992
QB$_1$.  {\it Nature} {\bf 362}, 730--730.

\medskip \noindent Jewitt, D., J. X.~Luu, and C.~Trujillo 1998. Large Kuiper Belt
objects: The Mauna Kea 8K CCD Survey. {\it Astron. J.} {\bf 115}, 2125--2135.

\medskip \noindent Levison, H. F., and M. J.~Duncan 1997. From the Kuiper Belt
to Jupiter-family comets: The spatial distribution of ecliptic comets.
{\it Icarus} {\bf 127}, 13--32.

\medskip \noindent Lissauer, J. J., and G. R.~Stewart 1993. Growth of planets from
planetesimals. In {\it Protostars and Planets III} (E. H.~Levy and J. I.~Lunine,
Eds.), pp. 1061--1088. Univ. of Arizona Press, Tucson.

\medskip \noindent Love, S. G., and T. J.~Ahrens 1996. Catastrophic impacts on
gravity dominated asteroids. {\it Icarus} {\bf 124}, 141--155.

\medskip \noindent Luu, J., and D.~Jewitt 1996. Color diversity among the Centaurs and
Kuiper Belt objects. {\it Astron. J.} {\bf 112}, 2310--2318.

\medskip \noindent Melosh, H. J., and E. V.~Ryan 1997.  Asteroids:  Shattered but not
dispersed.  {\it Icarus} {\bf 129}, 562--564.

\medskip \noindent Ryan, E. V., D. R.~Davis, and I.~Giblin 1999.  A laboratory impact
study of simulated Edgeworth-Kuiper Belt objects. {\it Icarus}, in press.

\medskip \noindent Ryan, E. V., and H. J.~Melosh 1998. Impact fragmentation: From the laboratory to asteroids. {\it Icarus} {\bf 133}, 1--24.

\medskip \noindent Stern, S. A. 1988. Collisions in the Oort Cloud.
{\it Icarus} {\bf 73}, 499--507.

\medskip \noindent Stern, S. A. 1995. Collisional time scales in the Kuiper
disk and their implications. {\it Astron. J.} {\bf 110}, 856--868.

\medskip \noindent Stern, S. A. 1996. Signatures of collisions in the Kuiper
disk. {\it Astron. Astrophys.} {\bf 310}, 999--1010.

\medskip \noindent Stern, S. A., and J. E.~Colwell 1997. Collisional erosion in
the primordial Edgeworth-Kuiper belt and the generation of the 30--50 AU
Kuiper gap. {\it Astrophys. J.} {\bf 490}, 879--882.

\medskip \noindent Stern, S. A., J. W.~Parker, M. J.~Duncan, J. C.~Snowdall, Jr., and H.
F.~Levison 1994.  Dynamical and observational constraints on satellites in the inner
Pluto-Charon system.  {\it Icarus} {\bf 108}, 234--242.

\medskip \noindent Terrile, R. J., S. A.~Stern, R. L.~Staehle, S. C.~Brewster, J. B.~Carraway,
P. K.~Henry, H.~Price, and S. S.~Weinstein 1997.
Spacecraft missions to Pluto and Charon. In {\it Pluto and Charon} (S. A.~Stern
and D. J.~Tholen, Eds.), pp. 103--124. Univ. of Arizona Press, Tucson.

\medskip \noindent Trafton, L. M., D. M.~Hunten, K. J.~Zahnle, and R. L. McNutt, Jr. 1997.
Escape processes at Pluto and Charon. In {\it Pluto and Charon} (S. A.~Stern
and D. J.~Tholen, Eds.), pp. 475--522. Univ. of Arizona Press, Tucson.

\medskip \noindent Veverka, J., P.~Thomas, A.~Harch, B.~Clark, J. F.~Bell~III,
B.~Carcich, J.~Joseph, C.~Chapman, W.~Merline, M.~Robinson, M.~Malin,
L. A.~McFadden, S.~Murchie, S. E.~Hawkins~III, R.~Farquhar, N.~Izenberg, and
A.~Cheng 1997. NEAR's flyby of 253 Mathilde: Images of a C asteroid. {\it Science},
{\bf 278}, 2109--2114.

\medskip \noindent Ward, W. R. 1996. Planetary Accretion. In {\it Completing the Inventory
of the Solar System, ASP Conference Series, Vol. 107} (T. W.~Rettig and J.M~Hahn,
Eds.), pp. 337--361. Astronomical Society of the Pacific, San Francisco.

\medskip \noindent Weissman, P. R., and H. F.~Levison 1997. The population of
the trans-Neptunian region: The Pluto-Charon environment. In {\it Pluto and
Charon} (S. A.~Stern and D.J.~Tholen, Eds.), pp. 559--604. Univ. of Arizona
Press, Tucson.

\medskip \noindent Weissman, P. R., and S. A.~Stern 1994. The impactor flux in
the Pluto-Charon system. {\it Icarus} {\bf 111}, 378--386.

\medskip \noindent Wetherill, G. W., and G. R.~Stewart 1993. Formation of planetary
embryos: Effects of fragmentation, low relative velocity, and independent variation
of eccentricity and inclination. {\it Icarus} {\bf 106}, 190--209.

\vfill
\eject

\centerline {\bf Figure Captions}

\medskip \noindent {\bf Figure 1}.  The critical eccentricity ($e^*$) boundary between
erosional and accretional outcomes for collisions between Edgeworth-Kuiper Belt objects.  Critical
eccentricity boundaries are shown for both strong ($\rho=2{\rm\ g\ cm^{-3}\ and\ } S_o=3
\times 10^{6}{\rm\ erg\ g^{-1}) \ and\ weak\ (}\rho=0.5{\rm\ g\ cm^{-3}\ and\ }S_o=3 \times
10^{4}{\rm\ erg\ g^{-1})}$ targets at two representative heliocentric distances (30 and 50
AU). The collision strengths chosen for the strong and weak cases bound a wide
range of material properties and, we believe, the likely range of collision strength
parameters of Kuiper Belt Objects.

\medskip \noindent {\bf Figure 2}.  The total number of impacts onto 102~km and 1~km radius
EKOs over 3.5 Gyr, as a function of projectile radius.  The bend in the plotted curves at
impactor radii of 10 km is a reflection of the shape of the EKO size distribution adopted in
our study (Weissman and Levison 1997).

\medskip \noindent {\bf Figure 3}.  Collisional disruption lifetime for EKOs as a function of
target size, calculated at 35 and 45 AU with $\langle e \rangle = 0.2048$.  Critical specific
energies for catastrophic disruption, $Q^*_D$, were assumed to be in the mid-range of those
from the published scaling laws referred to in the text.

\medskip \noindent {\bf Figure 4}.  Fraction of the surface area of 102~km and 1~km radius
EKOs cratered during 3.5 Gyr, as a function of projectile radius.

\medskip \noindent {\bf Figure 5}.  Kinetic energy partitioned into impact vaporization of
the surfaces of 102~km and 1~km EKOs during 3.5 Gyr, as a function of impactor radius.

\medskip \noindent {\bf Figure 6}.  Total number of impacts onto Pluto and Charon over 3.5
Gyr, as a function of impactor size.

\medskip \noindent {\bf Figure 7}.  Same as Fig.  1, except for the population of Centaur
objects.

\medskip \noindent {\bf Figure 8}.  Total number of impacts onto a roughly Chiron-size
Centaur of a typical dynamical lifetime in the giant planet region of 50 Myr, as a function
of impactor size.

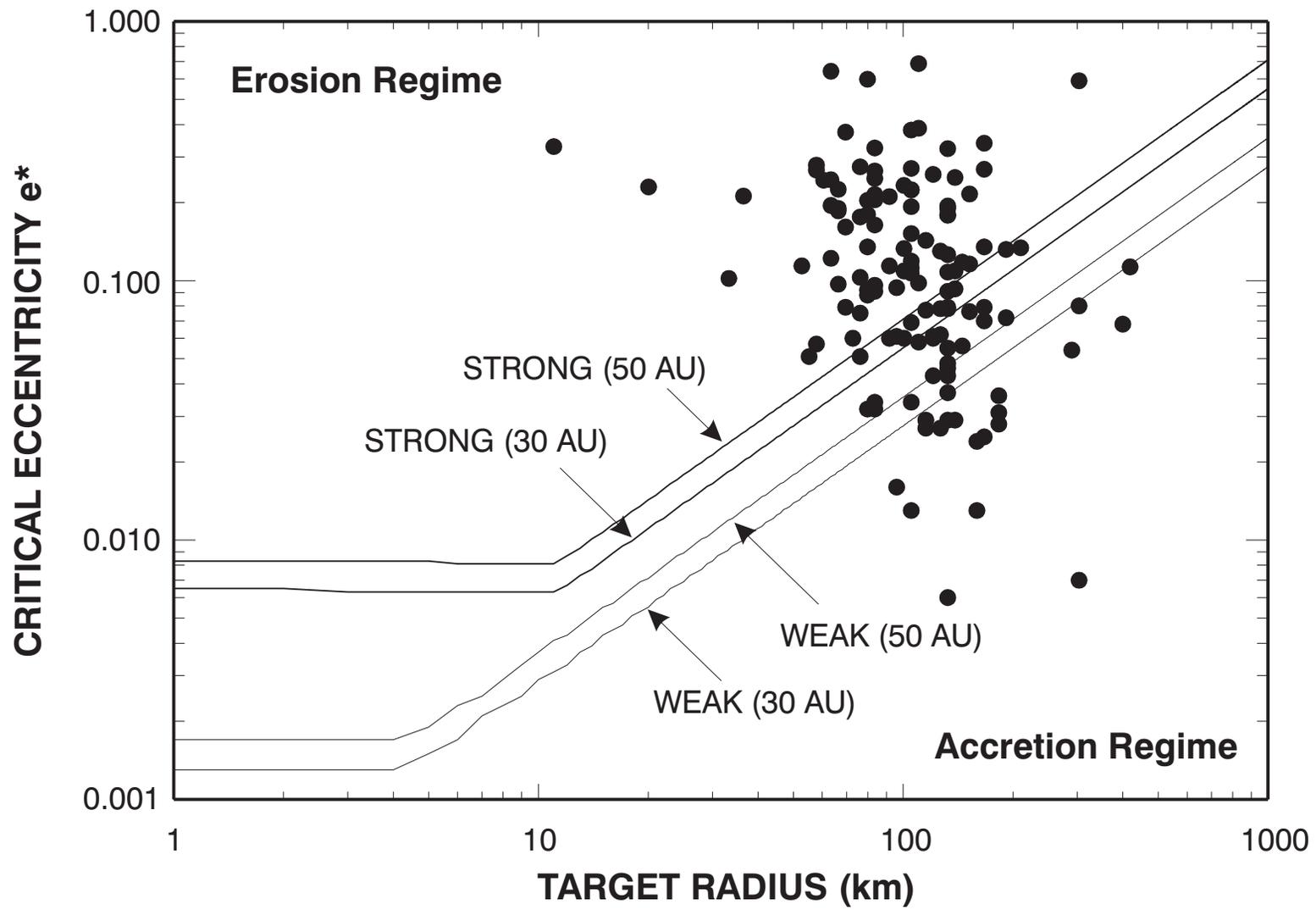
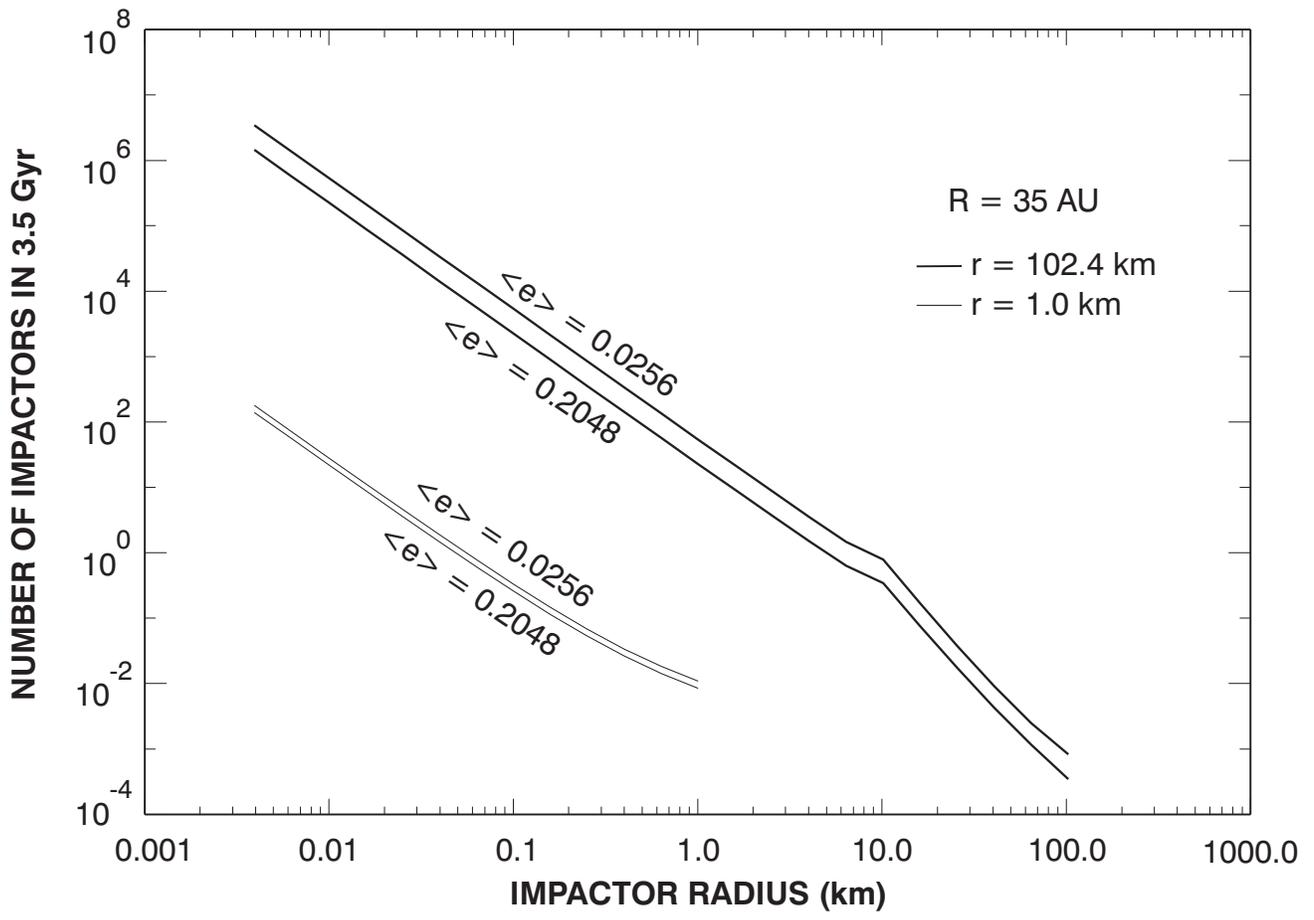
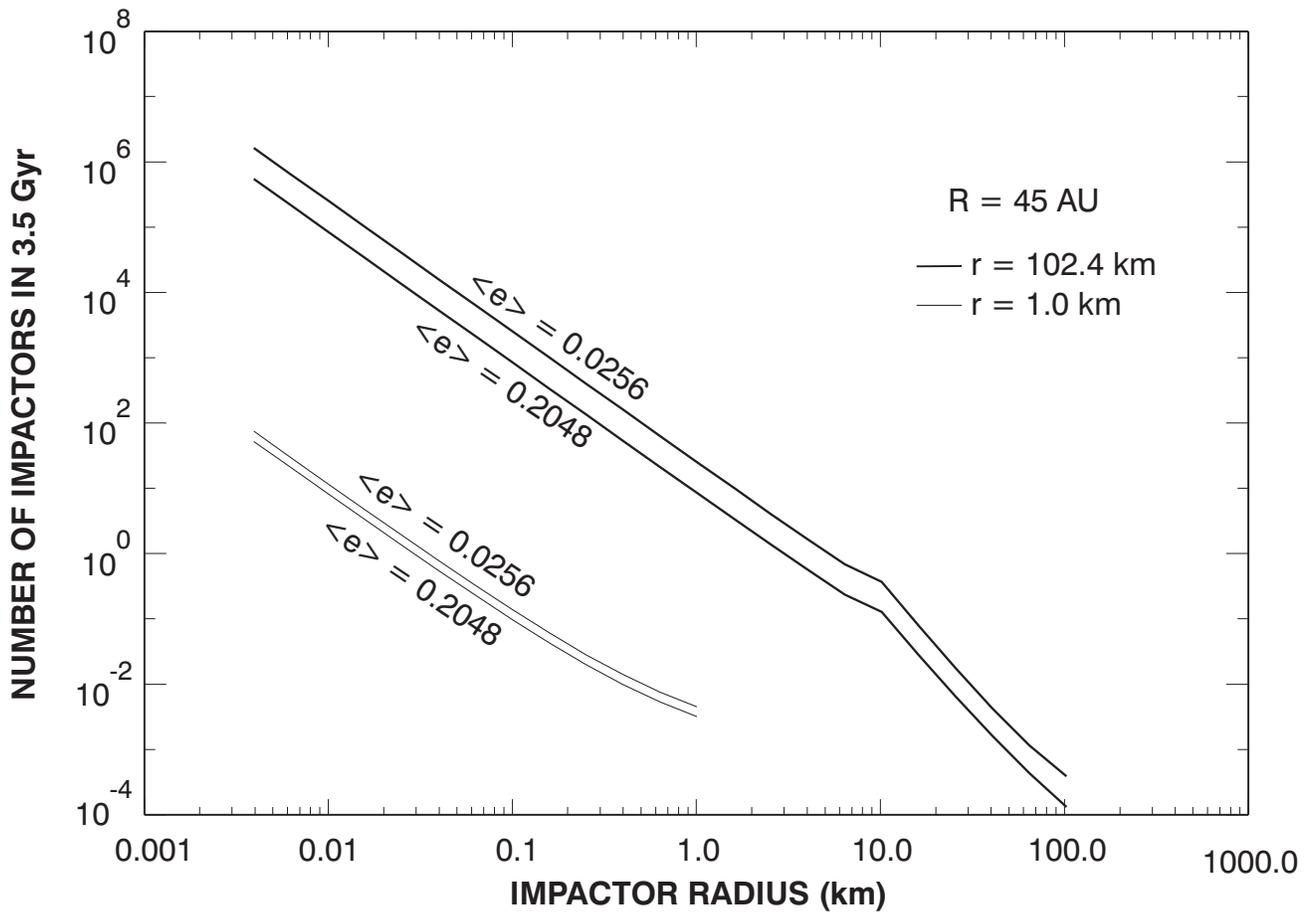
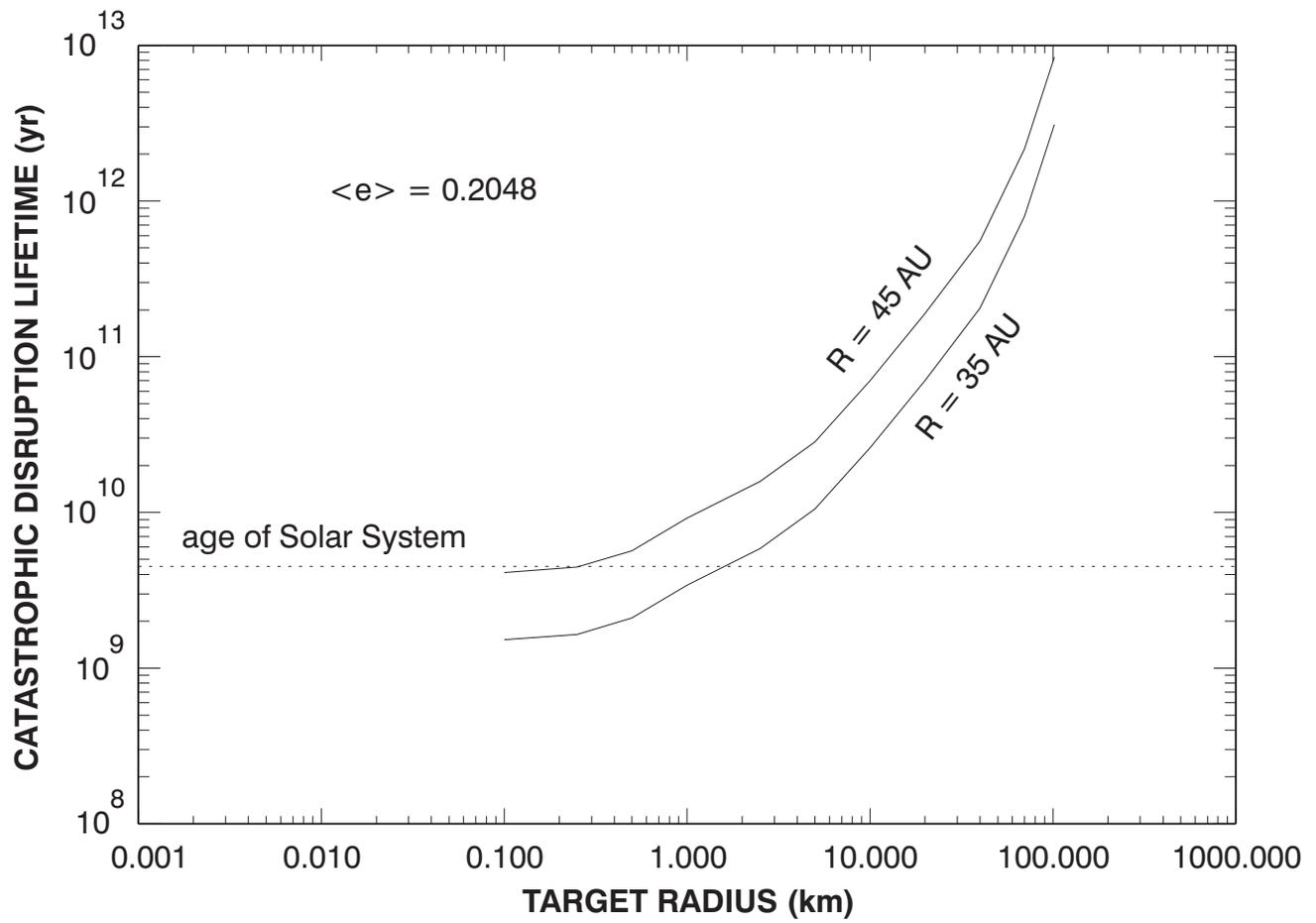
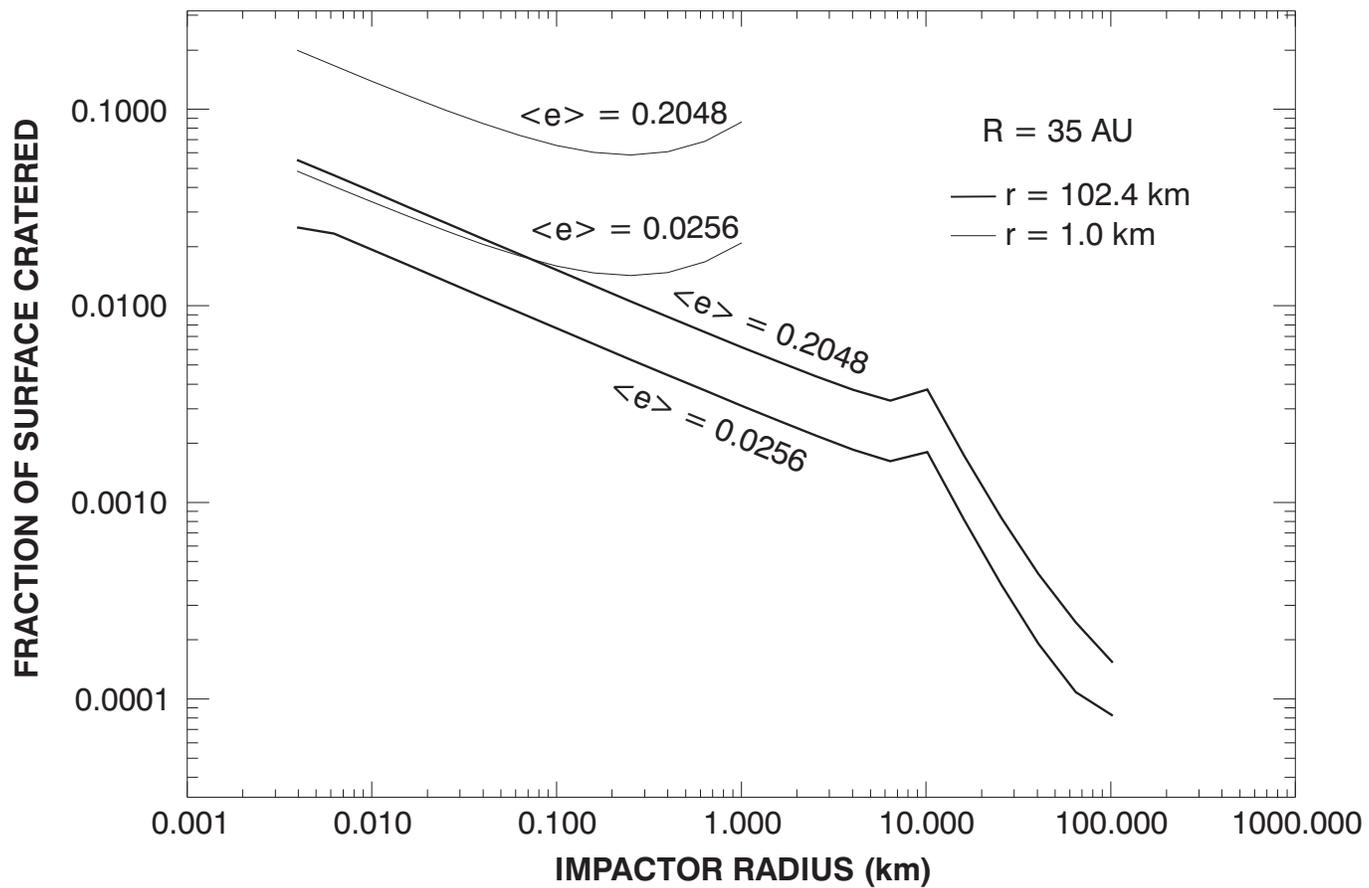
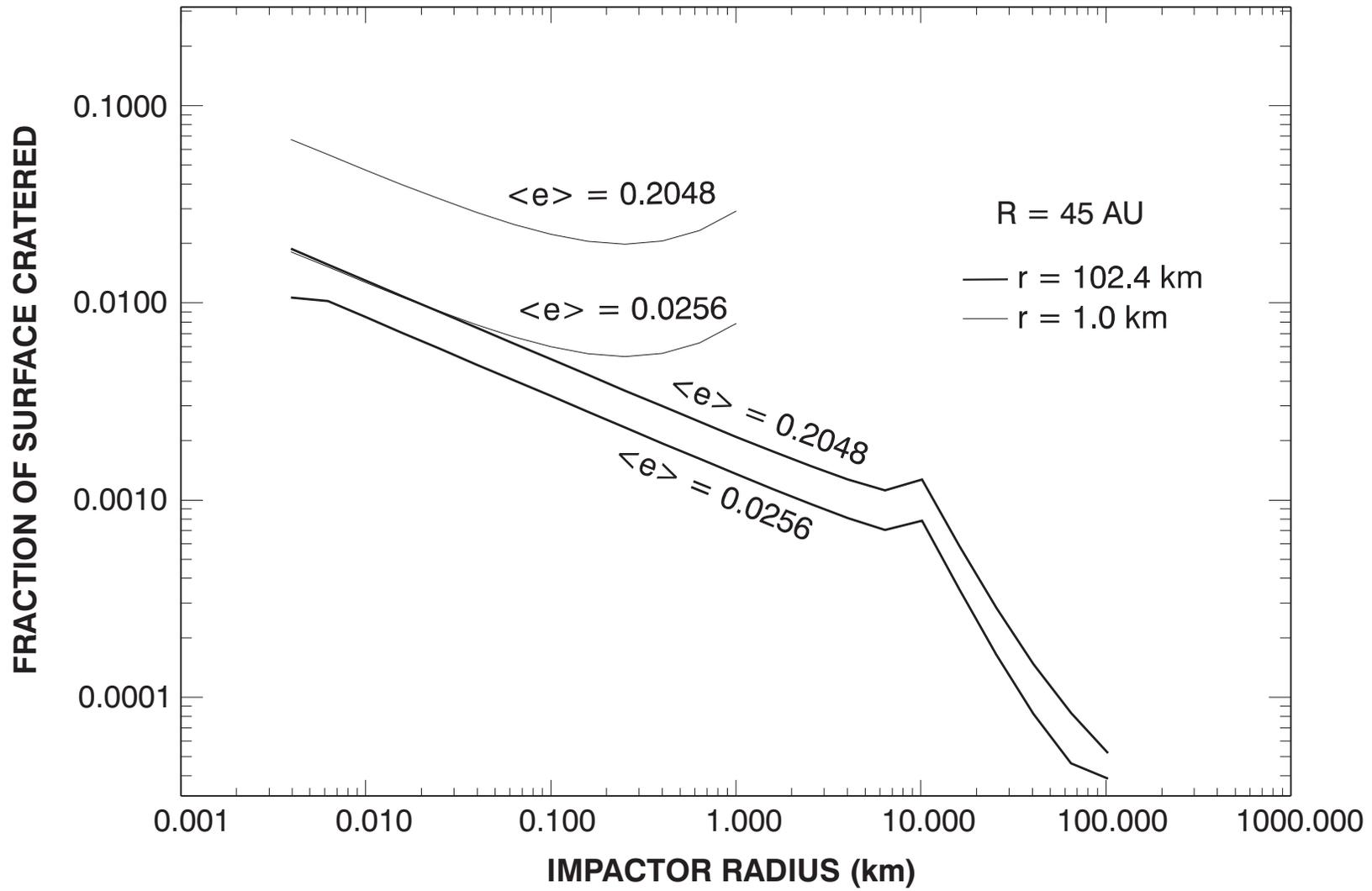
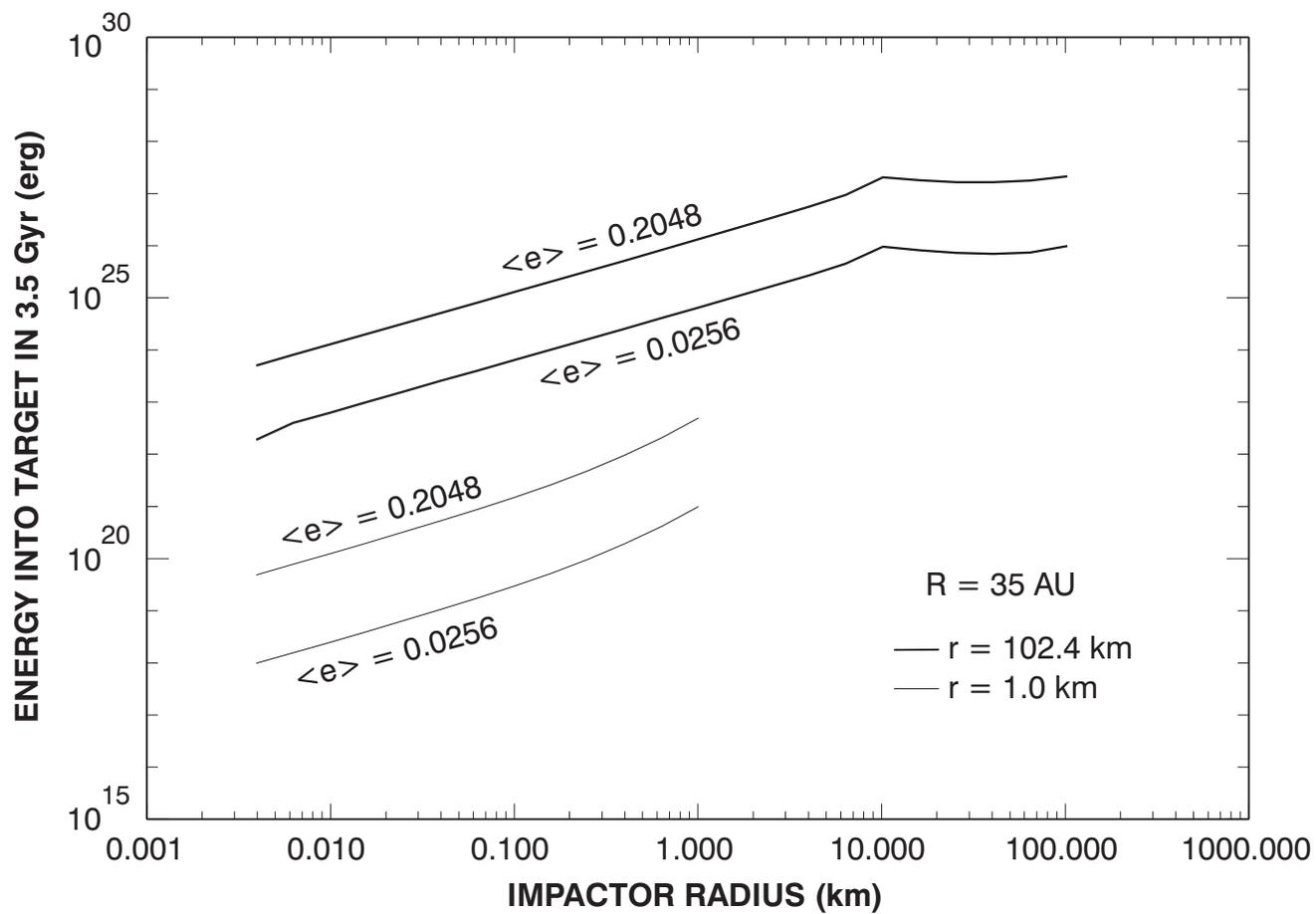
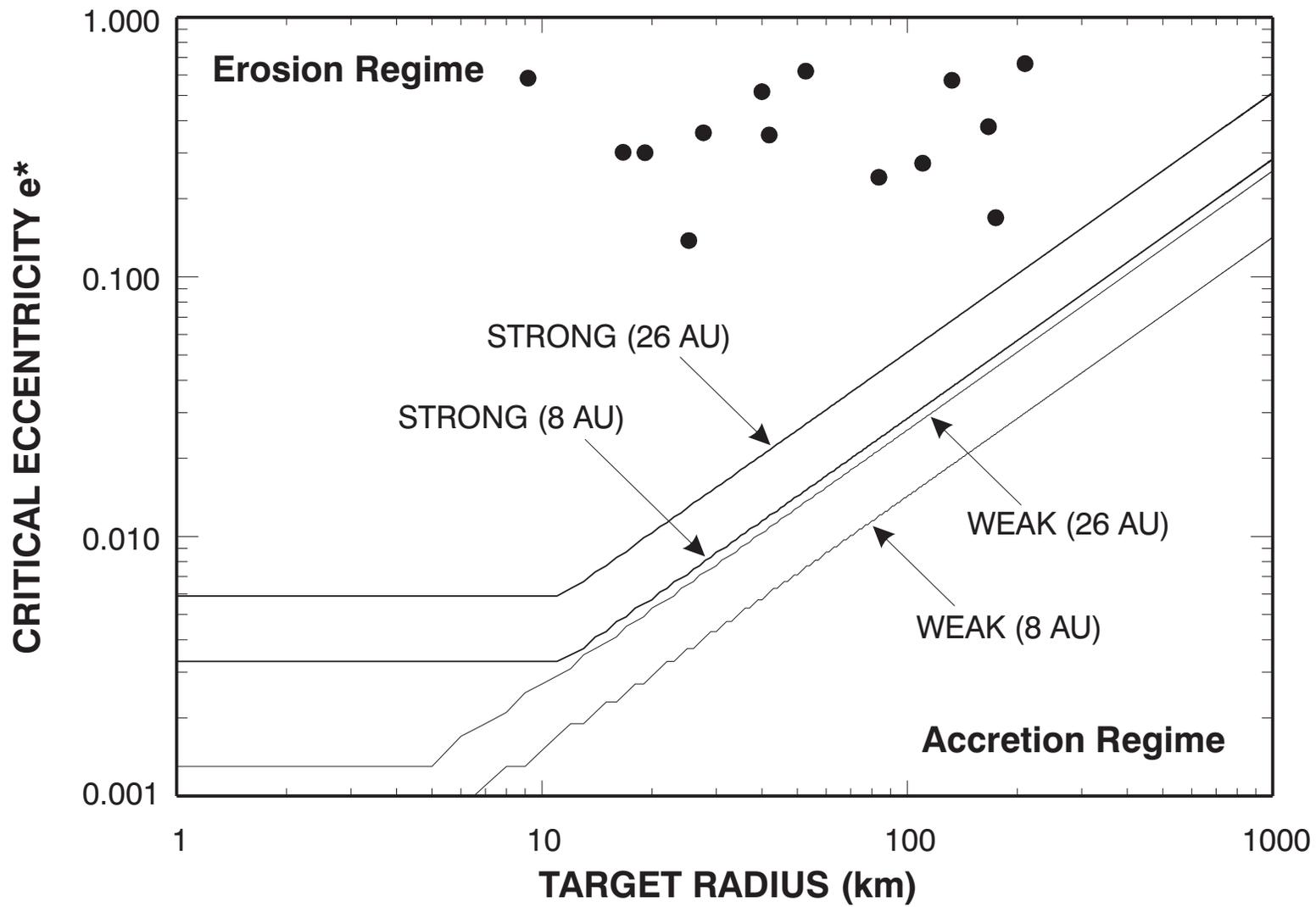
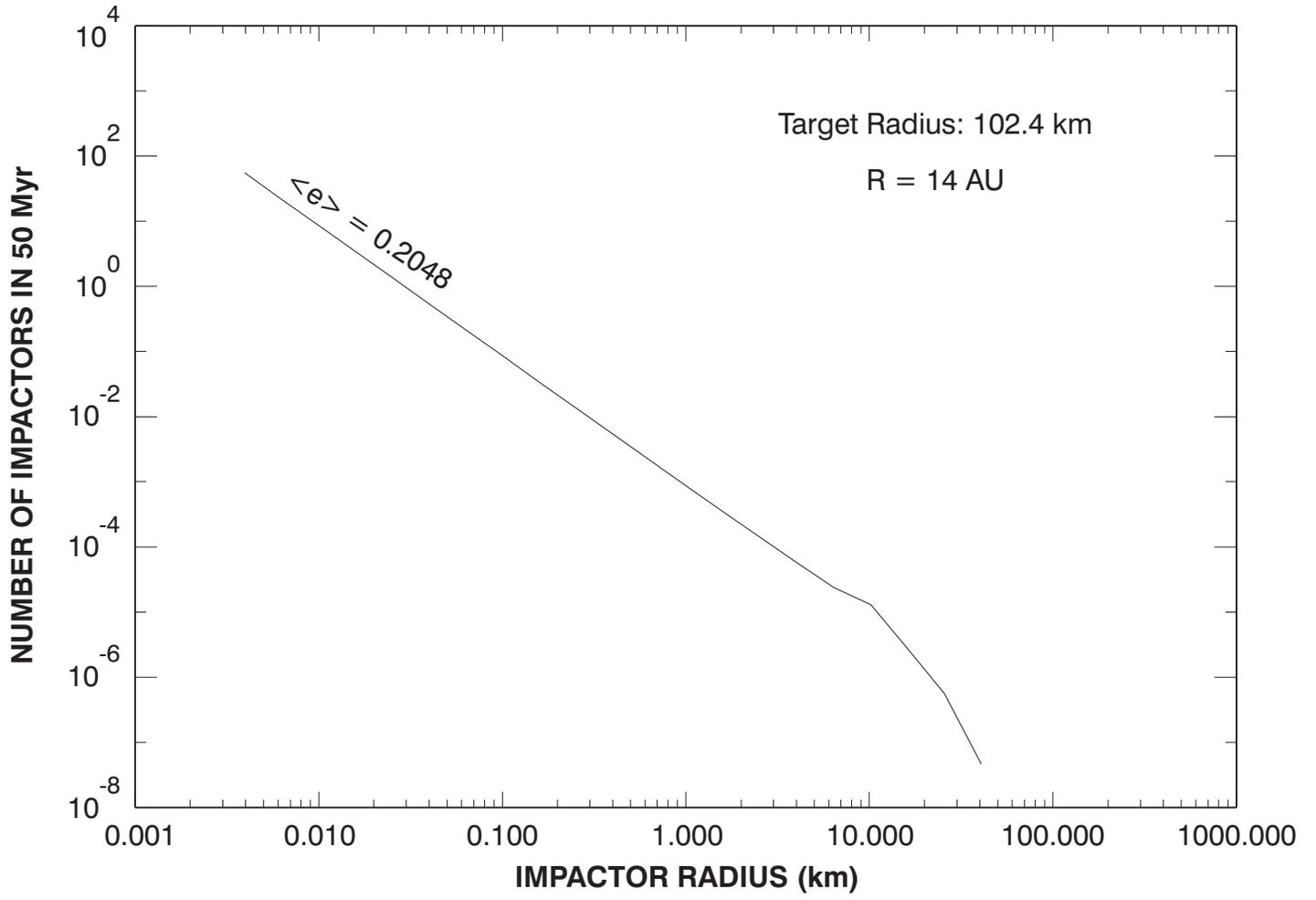
\bye